# A model driven method for promoting reuse in SOA-solutions by managing variability

**Boutaina Chakir [1], Mounia Fredj [2] and Mahmoud Nassar[3]**

**[1] AlQualsadi team, ENSIAS, Md V Souissi University**
**Rabat, Morocco**
*chakir.boutaina09@gmail.com*

**[2] AlQualsadi team, ENSIAS, Md V Souissi University**
**Rabat, Morocco**
*fredj@ensias.ma*

**[3] Models and Systems engineering team, ENSIAS, Md V Souissi University**
**Rabat, Morocco**
*nassar@ensias.ma*

## Abstract

Service Oriented Architecture (SOA) is an architectural paradigm that describes how organizations, people and systems provide and use services to achieve their goals and enhance productivity. Moreover, with the evolution of SOA, the focus in software development has shifted from applications to reusable services. However, the reuse in SOA is more seen as composition of fine-grained services rather than reuse of services implementation to build new services with additional functionalities. This can have some performance repercussions. Hence, in this paper, we propose a model driven method for managing Web service's variability based on MDA (Model Driven Architecture) as a way to promote reuse. In fact, through MDA, the method enables the automation of Web service's realization regardless of the supported platforms. Moreover, we present a WSDL extension meta-model called VarWSDL which enhances Web services by variability notions.

## 1. Introduction

Service Oriented Architecture (SOA) is an architectural paradigm based on the encapsulation of application logic within services that interact via a common communication protocol [1]. The most known type of services is the Web service which relies on the internet protocols for communication, and also exchanges data formatted as XML documents [1].

While technology and standards, such as Web services, are important to achieve SOA, they are not sufficient on their own. Indeed, with the growing of SOA and the multiplicity of implementation platforms, the emphasis is more focused nowadays on the service modeling techniques. In fact, it aims to define the concepts, formalisms and notations needed to describe SOA solutions regardless of the technologies and standards behind.

Several preliminary methodologies for service modeling have been proposed. Nevertheless, they are still immature and have not deeply addressed many challenges. Among these challenges, we quote reuse of web services. In fact, with the continuing evolution of information systems, the demanding customer's requirements increase the cost involved in designing and generating variants of a Web service in an ad-hoc manner. Hence, we need mechanism to reuse service capabilities to support the variety of usage scenarios.

To respond to this challenge, some approaches chose to implement a multi-purpose and coarse-grained service that is returning a large amount of data to consumers, although that many of them only require a subset. In consequence, this imposes an excess in data exchange and processing requirements upon service consumers. Others chose to realize a fine-grainer service that can be composed to provide capabilities. But in this case, there can be some performance repercussions of having to later compose together numerous fine-grained services.

In this paper, we try to overcome these difficulties by developing Web service families that share a set of functionalities by identifying and managing service's variability. Indeed, the aim of Web service's variability is to provide one central technique for better supporting the reusability of services in different application scenarios and to simplify the service consumption by variability resolution mechanisms [2].

In our previous work [3]-[4], we have presented our model driven service modeling method, which allows the representation of variability by dividing modeling process into two sub-processes. The first one is for commonalities and the second one is for the variability aspects of a system. The proposed modeling method is based on the model driven architecture (MDA), which enables the automation of the realization of the applications by the mechanisms of the models transformations. This makes the development decoupled from the platforms and also, productivity, quality and reuse enhanced. Also, we detailed the VarSOAML profile, which extends the SOAML (Service Oriented Architecture Modeling Language) meta-model [5] with variability information and allows the modeling at the PIM (Platform Independent Model) layer of our method. In this paper, we focus on the realization aspect of our method which represents the MDA-PSM (Platform Specific Model) layer. For this purpose, we propose a meta-model called VarWSDL, used for the generation of three documents that represent the variable Web service definition: the WSDL (Web Services Description Language) contract file, the XML Schema definition file and the variability specification file.

The remainder of the paper is organized as follows: Section 2 presents some related work. In Section 3, we give an overview of our method and the meta-model VarSOAML proposed for the variability modeling at the PIM layer. Section 4 describes in one hand the variability meta-model at the PSM layer called VarWSDL and on the other hand the transformation process from the VarSOAML (PIM level) to the VarWSDL (PSM level) meta-model. A case study is presented in Section 5 to illustrate our approach. Finally, the conclusion is reported in Section 6.

## 2. Related work

Service modeling and designing approaches are an active area of research that aims to specify systems in a high level of abstraction regardless of implementation technologies [6]. Among the studied approaches, we distinguish those that use existing development processes (like XP and RUP), the ones that propose their own processes and others that use the MDA approach. In the first category, we include the SOUP method [7], which proposes two variants: one adopting RUP for initial SOA projects and the other adopting a mix of RUP and XP for the maintenance of existing SOA projects. In the second category, we mention the following methods: Thomas Erl's [1], IBM Service-Oriented Analysis and Design (SOAD) [8], IBM Service Oriented Modeling and Architecture (SOMA) [9] which act at the provider side and support the services analysis and design. In the third category, we distinguish between approaches using the SOAML language proposed by the OMG and others using their own profile or meta-model. For instance, Kenzi et al. [10] propose a development method based on MDA and leaning on a UML profile called VSoaml in order to develop adaptable services with final user. Johnston and Brown [11] propose an approach based on UML profile that allows the design of services according to multiples modeling views. Elvesæter et al. [12], propose an MDA method based in the new language SOAML. In [13], they extend SOAML into «ShaML» (Semantically-enabled Heterogeneous Service Architecture and Platforms Engineering) in order to support several important aspects, like semantics, business modeling.

By studying these approaches, we found that they do not take into account the reusability of services or just consider it by the mean of services composition, which is not sufficient. In order to overcome this lack, one important issue to investigate is the management of variability, known as the most important way to promote reusability. In fact, the increasing need of agility and reuse of SOA solutions makes the variability a prominent issue to manage. Thus, some works are beginning to address this aspect under two perspectives: i) variability for reuse, where we try to make explicit the variability in design artifacts and to reuse the service at the provider level, ii) variability for adaptation, which focuses on the customization of a SOA-solution according to the context of use at the customer level.

In this paper, we are interested by the second perspective, i.e. the representation of variability for the purpose of web service's reuse improvement. Among the work reviewed, we quote [14], which focuses on the taxonomy of variation points for Web Service Flows as a starting point for handling variability through services. The work of [15] uses feature diagrams for modeling variability concept of the Web services. Besides, the authors of [16] propose a framework for modeling service variability. Also, they give a general classification of variation point in SOA solutions: Workflow, Composition, Interface and Logic according to SOAD architecture's layers. The work of [17] proposes a traceability model that maps requirements to use cases, sequence diagrams, business processes and finally service specifications. All these approaches do not consider all the stages of services development lifecycle, but just focus on the representation of variability at only one level: requirements or analysis. To address this concern, the authors of [2] propose an end-to-end approach for variability modeling called (VOE), that consists of three stages : Variation-Oriented Analysis (VOA), which considers the analysis of the solution with respect to its static and changing parts; Variation-Oriented Design (VOD), where a variation model for the solution design is instantiated based on the results of VOA; and Variation-Oriented Implementation (VOI), which consists of the implementation of variability based on the results of VOD. However, this approach is orthogonal to service design, not detailed and immature.

By reviewing these works, we found the absence of a comprehensive approach for service modeling that allows web services reuse by managing variability at different stages of the lifecycle development process. Furthermore, with the increasing number of platforms supporting SOA, it is interesting to use MDA in development approaches, because it makes the development decoupled from the platforms. This argues our approach for proposing a method for developing Web services; taking into account the variability management and following the MDA approach in order to promote reuse.

# 3. Multi-view model driven method for services modeling

## 3.1 Overview of the method

As we mentioned above, the proposed method described in Figure 1 is based on MDA. It is composed of three abstraction layers: the CIM (Computation Independent model) which describes the system's requirements, the PIM which specifies the system independently of any platform and the PSM which contains the services models related to a specific platform. In addition, in order to adopt the separation of concerns in our method, we divide our PIM layer into several modeling views [4]:

- Business view: contains the various business process models of the system, and the entity diagrams that capture the semantics of the solution.

- Service view: allows the identification of services. It contains the architecture, capability diagrams and contract diagrams.

- Functional view: contains service interface diagrams and messages exchanged among services.

- Interaction view: allows the representation of service compositions. Among the diagrams used in this view, we mention the service choreography diagram that models the interactions among services.

- Non-functional view: represents the non-functional properties applied to services.

Furthermore, the method is composed of two sub-processes:

- A base process composed of the following activities: service analysis and design, service realization and service implementation.

- A variability process which the main activities are: variability analysis and design, variability realization and variability implementation.

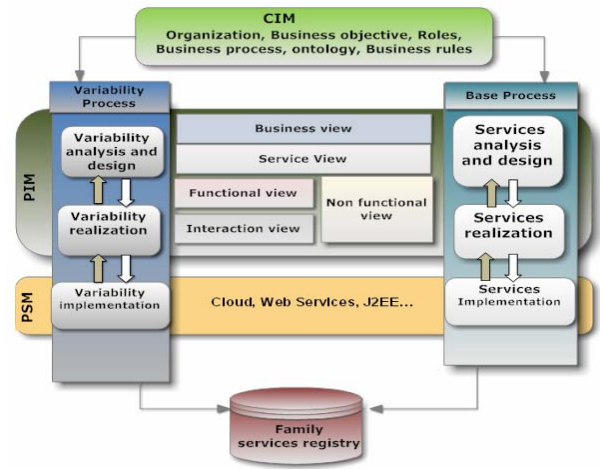

Fig. 1 The proposed modeling method.

In order to represent our models at the PIM layer, we propose a meta-model called VarSOAML presented in the next section.

## 3.2 Variability modeling at the PIM layer

### 3.2.1 VarSOAML meta-model

VarSOAML is a meta-model that results of merging two meta-models, one for services modeling and another one for representing services variability.

For services modeling, we adopt the SOAML language, because it is the first standard language proposed by the OMG for modeling SOA solutions. For variability modeling, we propose a ServiceVariability meta-model package containing variability elements (cf. Figure 2).

The main element of ServiceVariability meta-model is VariableElement that can take the following forms:

- VariableOperation: expresses a variant for the variation point representing a service operation.

- VariableMessage: represents a variant at the SOAML *MessageType*.

- VariableType: represents a variant associated to a *DataType*. It designates two elements:

  - ComplexVariableType: used for a *DataType*, that contains attributes.
  - SimpleVariableType: applied for a *DataType* that does not contain attributes.

To be able to use our language in an UML-based tool, it is mandatory to define an UML-Profile deriving our concepts to meta-classes defined by UML superstructure. Regarding the SOAML specification, the OMG group provides both a meta-model and a UML profile for the specification and design of services within a service-oriented architecture. For our ServiceVariability meta-model, we have introduced in [3], the equivalent stereotypes and the associated meta-classes.

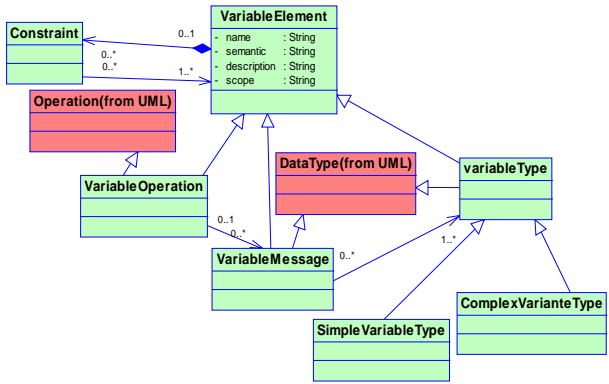

Fig. 2: The ServiceVariability meta-model package.

### 3.2.2 Application of the meta-model

In order to model variability of services, we use the following diagrams: interface diagram holding service interface, message diagram containing exchanged messages and interaction diagram representing the communication protocol of partners [3].

For interface diagram (cf. Figure 3), we use the element *VariableOperation*, which expresses a variable element associated to an operation. That is to say that this element is optional and according to user's requirements, it could not appear in the final interface.

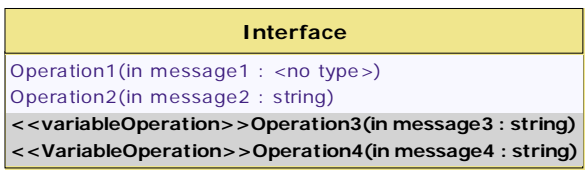

Fig. 3 Application of VariableOperation.

Regarding message diagram, we use three stereotypes to express the variation, depending on the type of variation point: VariableMessage, ComplexVariableType, SimpleVariableType. The Figure 4 illustrates the different types of the variability.

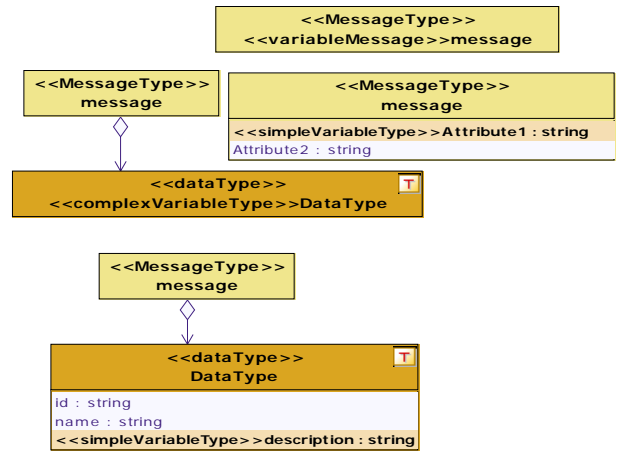

Fig. 4 Variability at message level.

In addition, for the interaction diagrams, we use only the "opt" operator (cf. Figure 5) defined in UML 2 for expressing optional fragment.

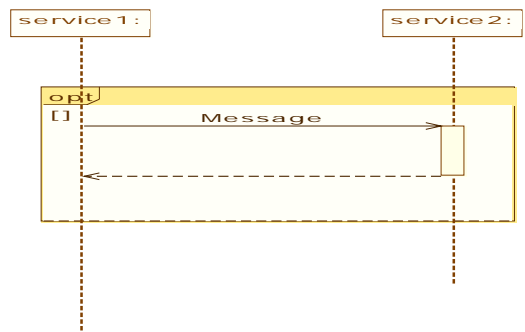

Fig. 5 Variability in the interaction diagram.

In the next section, we detail our representation of variability at the PSM layer.

## 4. Web services reusability

### 4.1 Variability modeling at the PSM layer

In our approach, we use the VarSOAML meta-model at the MDA-PIM layer for modeling services artifacts and the VarWSDL meta-model for the PSM layer. The PSM layer is composed of three files: the XML Schema file containing information description, the WSDL file describing service interface and the variability specification file that specifies the different variable elements of a service (cf. Figure 6).

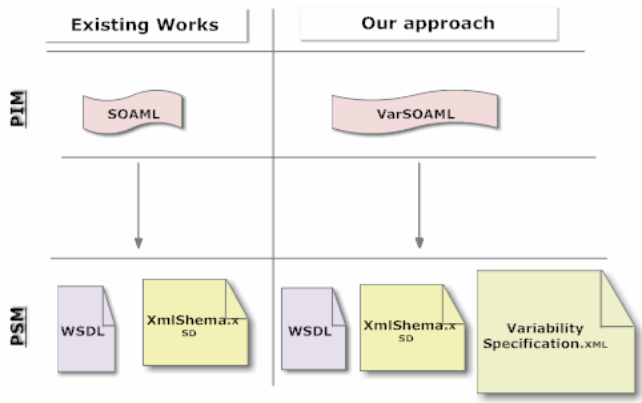

Fig. 6 Representation of services at the PSM layer.

In order to automatically generate these documents with our method following the MDA approach, we propose a VarWSDL meta-model. It is the result of the merging of three meta-models (cf. Figure 7):

- WSDL meta-model: inspired of the work of [18]. It contains elements that describe the WSDL specification. Among the elements of the meta-model, we list:

  - Definition: it is the parent element of each service description document. It contains all the other parts of the service: Import, Type, Message, PortType, Binding and Service.
  - Import: it specifies the location in which a namespace used within WSDL documents is established.
  - Types: it provides data type definitions used to describe the messages exchanged.
  - Message: it represents an abstract definition of the data being transmitted. A message is composed of a set of *Part* elements.
  - PortType: it is a set of abstract operations. Each operation refers to an input message, output message and fault message.
  - Binding: it assigns a communication protocol and data format specifications to the operations and messages defined by a particular PortType.
  - Port: it specifies an address for a binding.
  - Service: it provides a physical address at which the service can be accessed, which is used to aggregate a set of related ports.

- XSD meta-model: contains elements for the construction of the XMLShema document that describes the types used in the WSDL document. Using the XMLShema specification presented in [19], the meta-model is mainly composed of the following elements:

  - Schema: it is the parent element that contains all the other elements of the document.
  - Element: it represents a simple XMLSchema element, which is an XML element that can contain only text.
  - SimpleType: it defines a simple type and specifies the constraints and information about the values of attributes or text-only elements.
  - ComplexType: it describes a complex type element that is an XML element that contains other elements or attributes.
  - XsdPackageImport: this element is used to add multiple schemas with different target namespace to a document.

- ServiceVariability meta-model: contains elements for the generation of the variability specification file that represents the realization of the services variability (cf. Figure 8). Among the elements of the meta-model, we distinguish:

  - VariabilitySpecification: it is the parent element of the variability specification file. It contains the different variableElements of the service.
  - VariableElement: it describes a variation point of a service. Among its attributes, we specify *scope* that takes the values: optional, alternative or required.
  - Constraint: it represents dependency between variable elements.
  - VariableType: it can be ComplexVariabletype, which means variability in the whole elements of an XmlSchema Complextype. It can be also a SimpleVariableType, which concerns variability at the level of an XmlSchema element.
  - VariableOperation: it describes variability at the level of a service operation.
  - VariableMessage: it expresses variability at the level of service message. It is also, related to a *Part (element of the WSDL message)* which can be variable or not.

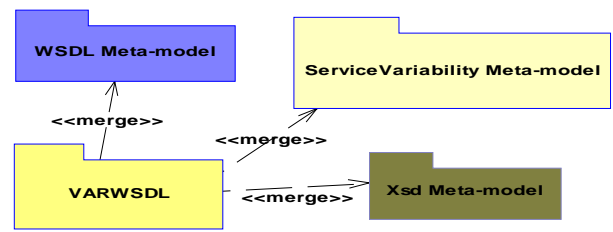

Fig. 7 VarWSDL meta-model.

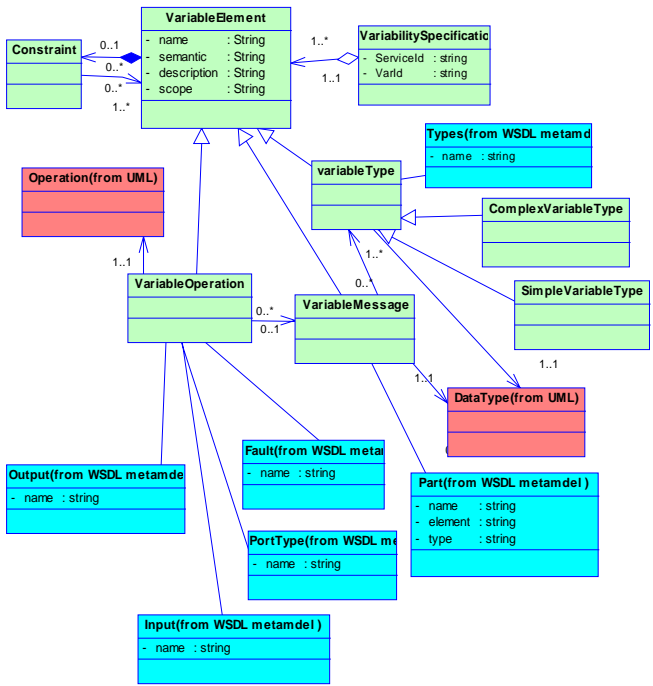

Fig. 8 Elements of the ServiceVariability meta-model.

Table 1: Mapping between VarSOAML and VarWSDL meta-models.

| VarSOAML meta-model elements | VarWSDL meta-model elements | Transformation rules |
|---|---|---|
| ServiceInterface | PortType, Binding, Types , Port, Definition | ServiceInterface2WSDL |
| Participant | Service | Participant2Service |
| Operation | PortypeOperation, BindingOperation, Input, Ouput, Fault | Operation2Operation |
| MessageType | Message, ComplexType | MessageType2Message |
| Attribute | Element | Attribute2Element |
| Datatype | ComplexType, Part | DataType2Type |
| UML primitive type | SimpleType | PType2SimpleType |
| VariableOperation (from VarSOAML: ServiceVariability) | VariableOperation (from VarWSDL: ServiceVariability) | VariableOperation2VariableOperation |
| VariableMessage (from VarSOAML: ServiceVariability) | VariableMessage (from VarWSDL: ServiceVariability) | VariableMessage2VariableMessage |
| SimpleVariableType (from VarSOAML: ServiceVariability) | SimpleVariableType (from VarWSDL: ServiceVariability) | SimpleVariableType2SimpleVariableType |
| ComplexVariableType (from VarSOAML: ServiceVariability) | ComplexVariableType (from VarWSDL: ServiceVariability) | ComplexVariableType2ComplexVariableType |

In order to proceed to the MDA transformation between the PIM and the PSM layer, we need to define the mapping between the elements of the source and the target meta-model, which will be presented in the next section.

## 4.2 Mapping between VarSOAML and VarWSDL meta-models

The first step we perform to transform the VarSOAML into VarWSDL is to identify the mappings. The mapping specification enables the determination of the equivalent elements between the source and the target meta-model elements. For this purpose, we implement several simple examples using both VarSOAML and VarWSDL.

In table 1, we expose the mapping between the two meta-models. The first column contains the elements of the VarSoaML source meta-model. The second one presents its equivalent elements from VarWSDL target meta-model and the third column presents the associated transformation rule that accomplishes the transformation of elements of source meta-model into elements of VarWSDL meta-model.

The second step of the transformation process consists of the definition of the transformation rules using ATL (Atlas Transformation Language:http://www.eclipse.org/m2m/atl). As illustrated in the table 1, we have defined a set of transformation rules. For instance, we present, in particular, the transformation rule ServiceInterface2WSDL. Such transformation rule creates an instance of a *PortType*, *Binding* and *Service* from the ServiceInterface element. The characteristics of *PortType*, *Binding*, and *Service* are assigned with the characteristics of the ServiceInterface. Thus, for the binding element, the name is assigned with the name of the ServiceInterface concatenated with the word 'Binding', *boperations* is filled by the collection of the operations of the services interfaces that are not a variableOperation element. The transport attribute is initialized with *http://schemas.xmlsoap.org/soap/http* and the style with 'rpc'. For the PortType element, the name is assigned with the name of the ServiceInterface concatenated with the word 'PortType' also *operations* is filled by the collection of the operations of the services interfaces that are not a *variableOperation* element. The ATL code below illustrates this transformation rule.

```
rule ServiceInterface2WSDL {

from inter : VarSOAML ! ServiceInterface
to bd : VarWSDL!Binding (
    name <- inter.name + 'Binding',
    type <- pType,
    transport <- 'http://schemas.xmlsoap.org/soap/http',
    style <- 'rpc',
    boperations <- inter.feature -> select
    (e|e.oclIsTypeOf(VarSOAML!Operation)) ->
    select(e| not e. oclIsTypeOf(VarSOAML!VariableOperation))
-> collect (e |thisModule.resolveTemp (e,' Operation2Operationb '))),

    pType: VarWSDL!PortType(
        name <- inter.name + 'PortType',
        operations <- inter.feature ->select
        (e|e.oclIsTypeOf(VarSOAML!Operation)) -> select(e| not e.
        oclIsTypeOf(VarSOAML!VariableOperation))
-> collect (e |thisModule.resolveTemp (e,' peration2Operationp '))),
    …
}
```

Fig. 9 ServiceInterface2WSDL transformation rule.

### 4.3 Model to code transformation for the VarWSDL meta-model

The generation of the specification files involves two steps as shown in (Figure 10). The first step, is the transformation from VarSOAML meta-model to VarWSDL meta-model handled in the last section. The second step is the generation of the specification file from the VarWSDL models. For this purpose, we define a set of helpers (functions) that are defined in the context of each element of the meta-model VarWSDL.

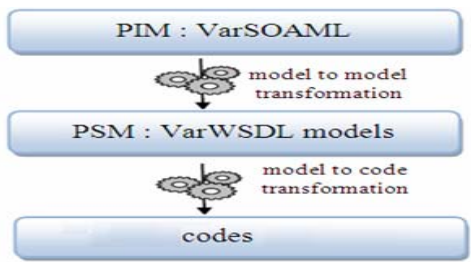

Fig. 10 Transformations from PIM to PSM and codes.

The figure 11 illustrates an example of ATL query used for this kind of transformation.

```
query var2code_query =
VarWSDL!VariabilitySpecification.allInstances() ->
collect(x|x.toString().writeTo('C:/SourceCode/'+
x.name.replaceAll(' .', '/') + '/' + x.name + '.xml'));
    -- Query Template

uses VarWSDL2code;
```

Fig. 11 VarWSDL code generation query.

The query var2code_query is responsible for the generation of the variability specification file related to the service (see Figure 17), the ATL operation allInstances() returns a set containing all the currently existing instances of the type VarWSDL!VariabilitySpecification. The operation collect () provided by ATL, returns a collection of elements which results in applying toString () helper (defined in the library VarWSDL2code) to each element of the source collection (represented by x).

## 5. Case study

To validate our approach, we use the application proposed by [20], describing "Supply Chain Management System" (SCMS). SCMS is the process of planning, implementing, and controlling the operations of the supply chain with the purpose to satisfy efficiently the customer requirements. In requirements specification stage, we identify the following participants: Client, Retailer, Warehouse, Shipper and Manufacturer. To illustrate the functional variability of the SCSM services, we consider that in the case of a foreigner client, some retailers may offer the possibility of checking whether the customer is eligible or not. By applying our method, the analysis of the SCMS system produces a set of models related to each modeling view:

- Business view: in this view, we represent all the business processes of the system by the UML activity diagram. These processes are used for the identification of services.
- Service view: the Figure 12 illustrates a simplified architecture diagram of our system. The SCMS system architecture binds the roles of participants: Client, Retailer, Shipper, Warehouse and Manufacturer. The participants participate to the following services: "purchasing service", "ship service", "ship status", "requesting goods service" and "manufacturing service".
- Functional view: SOAML offers the stereotype *MessageType* that constitutes the message diagram. The Figure 13 illustrates the message

diagram of our system. Also, in this view we construct the interface diagrams of the system (see Figure 14). The SOAML stereotype *ServiceInterface* defines the interface and the responsibilities of a participant to produce and consume a service. The interface with a name beginning with ~ indicates the conjugate interface (consumer level), using the operations offered by the interface of the service producer. Moreover, in variability sub-process, in order to represent the variability expressed in the requirements specification, at the level of the service proposed by the role "Retailer" called "purchasing service", we add the optional operation "isElligible". This operation takes as input the message "isElligibleRequest" and as output the integer (0, 1). The representation of this variability is shown in Figures 13-14.

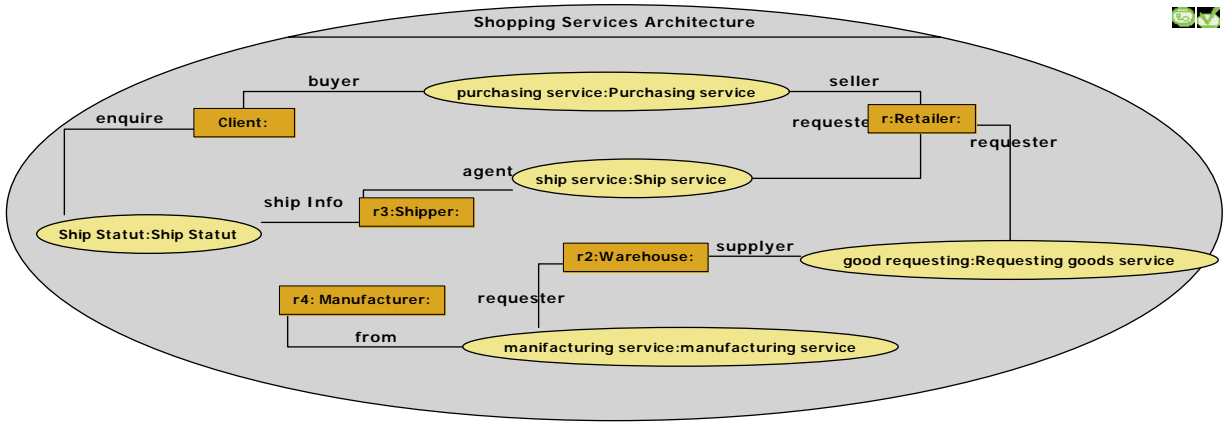

Fig. 12 Service view-architecture diagram of SCMS.

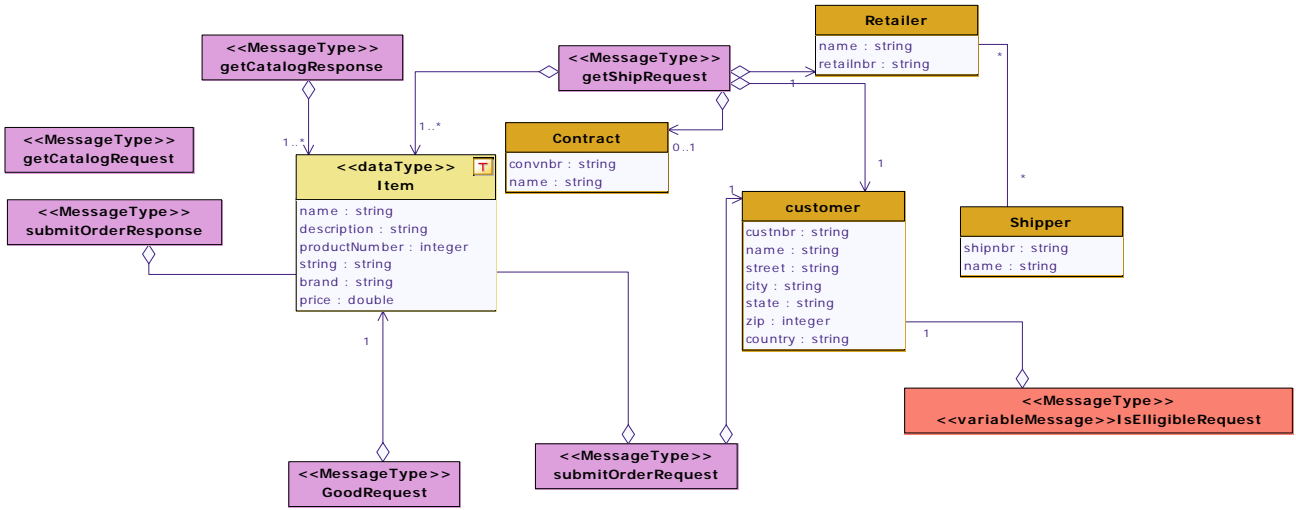

Fig. 13 Functional view-message diagram of SCMS.

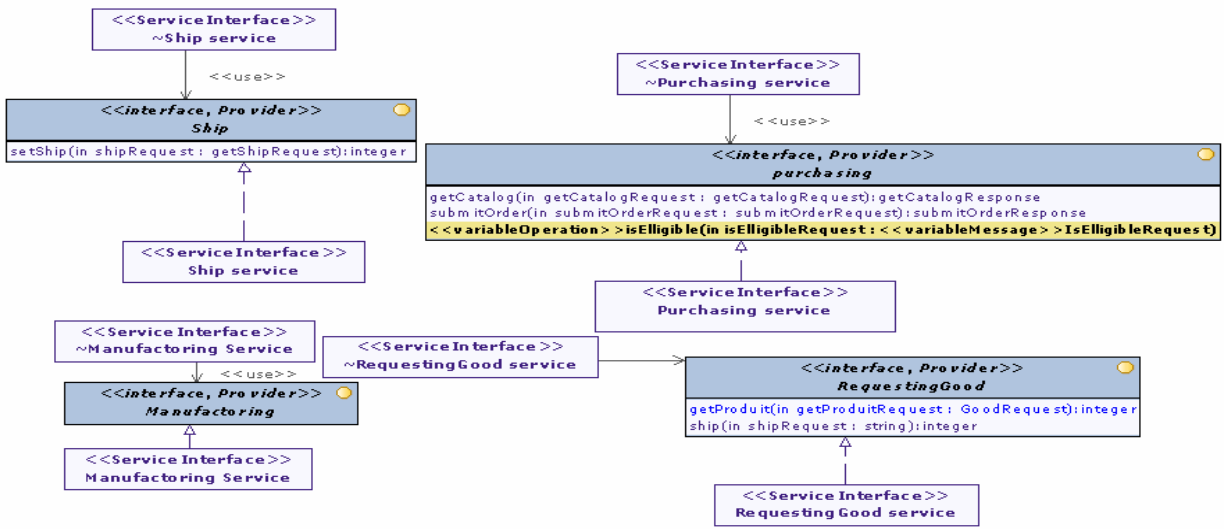

Fig. 14 Functional view-interface diagram of SCMS.

At the PSM level we generate service contract (.WSDL) and the specification of variability associated to the « purchasing » service.

```xml
<?xml version="1.0" encoding="UTF-8"?>
<wsdl:definitions

xmlns:xsd="http://www.w3.org/2001/XMLSchema"
xmlns:ps="http://scms/order.xsd"
xmlns:tns="http://scms/purchasingShema.wsdl"
targetNamespace="http://scms/purchasing.wsdl">
<wsdl:types>
  <xsd:schema><xsd:import
namespace="http://scms/purchasingShema.xsd"
schemaLocation="purchasingShema.xsd"/>
</xsd:schema>
</wsdl:types>
<wsdl:message name="getCatalogRequest"/>
<wsdl:message name="getCatalogResponse">
  <wsdl:part name="partItem"
type="ps:sequenceItem">
  </wsdl:part></wsdl:message>
…
<wsdl:portType name="purchasingPortType">
  <wsdl:operation name="getCatalog">

  <wsdl:input message="tns:getCatalogRequest"
name="getCatalogRequest"/>
  <wsdl:output message="tns:getCatalogResponse"
name="getCatalogResponse"/></wsdl:operation>
<wsdl:operation name="submitOrder">
  <wsdl:input message="tns:submitOrderRequest"
name="submitOrderRequest"/>
  <wsdl:output message="tns:submitOrderResponse"
name="submitOrderResponse"/> </wsdl:operation>
                    </wsdl:portType>
…
</wsdl:definitions>
```

Fig. 15 "purchasing" service contract.

```xml
<?xml version="1.0" encoding="UTF-8"?>
<xsd:schema
xmlns:xsd="http://www.w3.org/2001/XMLSchema"
elementFormDefault="qualified"
  xmlns:tns="http://scms/purchasingShema.xsd"
targetNamespace="http://scms/purchasingShema.xsd">
<xsd:complexType name="sequenceItem">
<xsd:sequence>
  <xsd:element name="elementItem" type="tns:item"
/>
</xsd:sequence>
</xsd:complexType>
<xsd:complexType name="item">
<xsd:sequence>
  <xsd:element name="name"    type="xsd:string"
minOccurs="1" maxOccurs="1"/>
  <xsd:element name="description"
type="xsd:string" minOccurs="1" maxOccurs="1"/>
  <xsd:element name="productNumber"
type="xsd:integer" minOccurs="1" maxOccurs="1"/>
  <xsd:element name="category" type="xsd:string"
minOccurs="1" maxOccurs="1"/>
  <xsd:element name="brand" type="xsd:string"
minOccurs="1" maxOccurs="1"/>
  <xsd:element name="price" type="xsd:decimal"
minOccurs="1" maxOccurs="1"/>
</xsd:sequence>
</xsd:complexType>
<xsd:complexType name="customer">
<xsd:sequence>
  <xsd:element name="custnbr"
type="tns:CustomerReferenceType" minOccurs="1"
maxOccurs="1"/>
  <xsd:element name="name"    type="xsd:string"
minOccurs="1" maxOccurs="1"/>…
</xsd:sequence>
</xsd:complexType>
…
</xsd:schema>
```

Fig. 16 XMLSchema associated to the "purchasing" service.

```xml
<?xml version="1.0" encoding="UTF-8"?>
<variability
xmlns:ps="http://localhost/test/purchasingShema.xs
d" xmlns:xsd="http://www.w3.org/2001/XMLSchema">
<service> 1111 </service>
<varId> 11 </varId>
<Operations>
<variableOperation>
<ID>1 </ID>
<NAME> isElligible</NAME>
<SCOPE>optional</SCOPE>

<BOUNDeLEMENT>isElligible</BOUNDeLEMENT>

<input message="1.2" name="isElligibleRequest"/>
<output message= "1.3"
name="isElligibleResponse"/>
</variableOperation >
</Operations>
<messages>
<variablemessage>
 <ID>1.2 </ID>
 <NAME> isElligibleRequest</NAME>
 <SCOPE> required </SCOPE>
 <BOUNDeLEMENT>isElligibleRequest</BOUNDeLEMENT>
    <part   name="partCustomer"
type="customerVariable" vp="s"/>
</variablemessage>
<variablemessage>
 <ID>1.3 </ID>
 <NAME> isElligibleResponse</NAME>
 <SCOPE> required </SCOPE>
 <BOUNDeLEMENT>isElligibleResponse</BOUNDeLEMENT>
    <part   name="partIsElligibleResponse"
type="xsd.integer" vp="no"/>

</variablemessage>
</messages>
<types>
<Simplevariabletype>
<xsd:complexType name="customerVariable">
<xsd:complexContent>
<xsd:extension base="ps:customer">
<xsd:sequence>
<xsd:element name="address" type="xsd:string" />
<xsd:element name="city" type="xsd:string"/>

<xsd:element name="country" type="xsd:string"/>
</xsd:sequence>
</xsd:extension>
</xsd:complexContent>
</xsd:complexType>
<ID/>
<NAME/>
<SCOPE> optional</SCOPE>
<BOUNDeLEMENT>customer</BOUNDeLEMENT>
</Simplevariabletype>
```

Fig. 17 Variability configuration file associated to the "purchasing" service.

## 6. Conclusion

With the evolution of SOA, especially its Web-based implementation (Web services), service modeling approaches become very important. This has raised several challenges such as web services reuse. Among the approaches that allow the software reuse, we include the variability management. To address this issue, we propose a model driven method based on MDA for developing Web services and allowing the variability representation in order to permit the web services reuse in multiple contexts. This method is supported by two meta-models for the PIM and the PSM layers, and by a transformation process to allow the automatic generation of the realization models.

Our future work will give the definitions of the meta-model for the implementation code and the associated transformation rules, in order to facilitate its automatic generation. Also, an adaptation tool will be implemented for the resolution of variability at the client s' side.

## References

[1] T. Erl, Service-Oriented Architecture: Concepts, Technology, and Design, Prentice Hall PTR, 2005.

[2] N.C. Narendra, K. Ponnalagu, B. Srivastava and G.S. Banavar, "Variation-oriented engineering (VOE): Enhancing reusability of SOA-based solutions", in IEEE International Conference on Services Computing, SCC'08, 2008, 1, pp. 257-264.

[3] B. Chakir and M. Fredj, "Towards a modelling method for managing variability in SOA", in Proceedings of the IADIS International Conference Information Systems 2011, Avila, Spain, 2011, pp. 296-300.

[4] B. Chakir and M. Fredj, "A model driven approach supporting multi-view services modeling and variability management", in Proceedings of the International Conference on Enterprise Information Systems (ICEIS), Beijing, China, June, 2011.

[5] SOAML, Service Oriented Architecture Modeling Language (SoaML)- Specification for the UML Profile and Metamodel for Services (UPMS). FTF Beta 2, Available in: http://www.omg.org/docs/ad/08-08-04.pdf, 2009.

[6] M. Papazoglou, P. Traverso, S. Dustdar and F. Leymann. "Service-Oriented Computing: State of the Art and Research Challenge", IEEE Computer Society, Vol. 40, No. 11, 2007, pp. 38-45.

[7] K. Mittal, Service Oriented Unified Process (SOUP), available in: http://www.Kunalmittal.com/ html/soup.shtml, 2006.

[8] O. Zimmermann, P. Krogdahl and C. Gee, "Elements of Service-Oriented Analysis and Design", IBM developerWorks, 2004.

[9] A. Arsanjani, Service-oriented modeling and architecture, IBM Corporation, available in http://www-128.ibm.com/developerworks/web services /library/ws-soa-design1/, 2004.

[10] A. Kenzi, B. El Asri, M. Nassarand and A. Kriouile, "A modeldriven framework for multiview service oriented system development", in Proceedings of the 7th ACS/IEEE International Conference on Computer Systems and Applications AICCSA'09, Rabat, Morocco, 2009.

[11] S. K Johnston. and A. Brown, " A Model-Driven Development Approach to Creating Service-Oriented Solutions", ICSOC 2006: 624-636.

[12] B. Elvesæter, C. Carrez, P . Mohagheghi., A.-J Berre, S. G. Johnsen, and A. Solberg, "Model-driven Service Engineering with SoaML", in Service Engineering -

European Research Results, Wien, Springer, 2011, pp. 25-54.

[13] SHAPE, Semantically-enabled Heterogeneous Service Architecture and Platforms Engineering, available in http://www.shape-project.eu, 2009.

[14] S. Segura, D. Benavides, A. Ruiz-Cort´es and P. Trinidad, "A taxonomy of variability in Web service flows", In first workshop on Service-Oriented Architecture and Software Product Lines, 2008.

[15] S. Robak and B. Franczyk, "Modeling Web services variability with feature diagrams", in: Revised Papers from the NODe 2002 Web and Database-RelatedWorkshops on Web, Web-Services, and Database Systems, Springer,Verlag, 2003, pp. 120-128.

[16] SH. Chang and SD. Kim, "A variability modeling method for adaptable services in service-oriented computing", Software Product Line Conference, 2007, pp. 261-268.

[17] N. Narendra, K. Ponnalagu, "Variation-Oriented Requirements Analysis (VORA)", IEEE Congress on Services, 2007, pp.159-166.

[18] A. Kenzi, B. El Asri, M. Nassar and A. Kriouile, "A modeldriven framework for multiview service oriented system development", Proceedings of the 7th ACS/IEEE International Conference on Computer Systems and Applications AICCSA'09, Rabat, Morocco, 2009.

[19] E. van der Vlist, XML Schema, O'Reilly Vlg. GmbH & Co, February 28, 2003.

[20] (WS-I), Web Services Interoperability Organization (WS-I) Web site, Available in: http://www.ws-i.org/,2007.

**Boutaina Chakir** is a Software Engineer graduated from ENSIAS (2004) (National Higher School for Computer Science and System analysis), holder of an Extended Higher Studies Diploma from ENSIAS (2007) and "Ph.D. candidate" at ENSIAS. Her research focuses on management variability in SOA. She is a project engineer at the Ministry of Economy and Finance of the Kingdom of Morocco since 2004.

**Mounia Fredj** is a PhD in Computer Sciences. Professor in ENSIAS, (National Higher School for Computer Science and System analysis), Rabat, Morocco; Ongoing research interests: Information System Engineering, components, patterns, MDA approach.

**Mahmoud Nassar** is a PhD in Computer Sciences. Professor in ENSIAS, (National Higher School for Computer Science and System analysis), Rabat, Morocco. Ongoing research interests: Object modeling of complex system based on view point (VUML method), MDA approach and context aware SOA.